\documentstyle[12pt,aaspp4]{article}
\begin{document}

\parindent=1.0cm

\title{The Asymptotic Giant Branch of NGC 205: The Characteristics of Carbon 
Stars and M Giants Identified From $JHK'$ Images}

\author{T. J. Davidge \footnote[1]{Visiting Astronomer, Canada-France-Hawaii
Telescope, which is operated by the National Research Council of Canada, 
the Centre National de la Recherche Scientifique, and the University of 
Hawaii}}

\affil{Canadian Gemini Office, Herzberg Institute of Astrophysics,
\\National Research Council of Canada, 5071 West Saanich Road, 
\\Victoria, B.C. Canada V9E 2E7\\ {\it email: tim.davidge@nrc-cnrc.gc.ca}}

\begin{abstract}

	$J, H,$ and $K'$ images are used to investigate the asymptotic giant 
branch (AGB) content of the Local Group dwarf elliptical galaxy NGC 
205. The AGB on the $(K, H-K)$ and $(K, J-K)$ color-magnitude diagrams 
consists of two sequences: a near-vertical plume of giants 
with spectral types K and M, and a red arm containing C stars. 
There are 320 C stars with M$_{bol} < -4.1$ and $J-K > 1.5$ 
within 2 arcmin of the nucleus. C stars account for 10\% of the 
integrated luminosity of AGB stars brighter than M$_{bol} = -3.75$ near the 
center of NGC 205, and this is in excellent agreement with what is measured in 
intermediate-age clusters in the LMC. The most luminous AGB star has M$_{bol} = 
-6.5$, although variability introduces an uncertainty of a few tenths 
of a magnitude when using this as an estimate of the AGB-tip brightness. 
Comparisons with models suggest that the brightest AGB stars formed 
within the past 0.1 Gyr, and that the previous episode of star formation 
occured a few tenths of a Gyr earlier. These results are consistent with star 
formation in NGC 205 being triggered by interactions with M31. These 
data also demonstrate that near-infrared imaging provides an efficient means of 
identifying C stars in nearby galaxies. The techniques used here to 
identify C stars and probe the AGB are well suited to studies of 
galaxies outside of the Local Group using data obtained 
with adaptive optics systems on large ground-based telescopes.

\end{abstract}

\keywords{galaxies: individual (NGC 205) -- galaxies: stellar content -- galaxies: dwarf}

\section{INTRODUCTION}

	The Local Group spiral galaxy M31 has a rich entourage of companions, 
including the 3 dwarf elliptical galaxies (dEs) NGC 147, NGC 185, and NGC 205, 
the compact elliptical galaxy M32, and a number of dwarf spheroidal galaxies 
(dSphs). NGC 205 is the brightest of the dEs, and may be the nearest example of 
a nucleated dE (Zinnecker \& Cannon 1986). With a distance modulus of 24.6 
(Saha, Hoessel, \& Krist 1992), NGC 205 is 100 kpc behind M31, and 
there are indications that NGC 205 has interacted with M31 and its 
companions in the past. Cepa \& Beckman (1988) point out that 
NGC 205 and M32 have similar orbital properties, and Ibata et al. (2001) find 
a tidal stream in the M31 halo that is aligned with M32 and NGC 205. 
Sato \& Sawa (1986) argue that NGC 205 may have warped the 
HI disk of M31, and the structural characteristics of NGC 205 show 
classic signatures of tidal interactions (Choi, Guhathakurta, \& Johnston 
2002). The ISM of NGC 205 is also smaller than expected 
given the rate of replenishment from stellar mass loss 
(Welch, Sage, \& Mitchell 1998), as expected if the 
gas and dust are periodically stripped away by tidal interactions.

	Previous studies of the resolved stellar content of NGC 205 have found 
stars spanning a range of ages. Stars evolving on the red 
giant branch (RGB), which has a color indicative of [Fe/H] $= -0.85$ 
and a width suggesting that $\sigma_{[Fe/H]} = 0.5$ dex 
(Mould, Kristian, \& da Costa 1984), are among 
the brightest members of the old stellar substrate. There is an extended AGB, 
which Richer, Crabtree, \& Pritchet (1984), Davidge (1992), and Lee (1996) find 
has a peak M$_{bol}$ between --5.5 and --6, indicating that NGC 205 formed 
stars during intermediate epochs; Richer et al. (1984) also identified 
7 C stars near the center of NGC 205, while Demers, Battinelli, \& 
Letarte (2003) have recently found 500 C stars 
scattered throughout the galaxy. It has long been known that 
there are young blue stars in the central regions of NGC 205 (e.g. Baade 
1951 and discussion therein), and many of these have since been found to 
be associations or clusters (Cappellari et al. 1999). While the nuclear 
regions of NGC 205 have a flat spectral-energy distribution (SED) 
in the UV, with UV-bright stars contributing 60\% of the flux between 1200 and 
2450\AA\ (Bertola et al. 1995), young stars likely account for 
less than 1\% of the total stellar mass (Wilcots et al. 1990).

	In the present study, deep $J, H,$ and $K'$ images are used 
to conduct the first investigation of the resolved stellar content 
of NGC 205 at wavelengths longward of $1\mu$m. The reddest, most extreme, 
AGB stars can be difficult to detect at visible wavelengths, and are 
more easily detected in the infrared. The resulting increased sensitivity to 
the cool stars that are the brightest members of old and intermediate age 
populations makes it easier to probe the star-forming history of the 
crowded central regions of the galaxy. The majority of bright C 
stars (i.e. those that are not `warm') also have near-infrared 
SEDs that differ from those of oxygen-rich M giants (e.g. Wood, 
Bessell, \& Paltoglou 1985; Hughes \& Wood 1990), so that broad-band 
infrared colors, which can be obtained from moderately short exposures, 
can be used to distinguish between these two types of objects.
 
\section{OBSERVATIONS, REDUCTIONS, AND PHOTOMETRIC MEASUREMENTS}

	The data were recorded on UT June 4 2001 with the CFHTIR imager, 
which was mounted at the Cassegrain focus of the 3.6 metre Canada 
France Hawaii Telescope. CFHTIR contains a $1024 
\times 1024$ Hg:Cd:Te array. Each pixel samples 0.21 arcsec on a 
side, so that a $3.6 \times 3.6$ arcmin field is imaged.
Data were recorded through $J, H,$ and $K'$ filters, 
with a total exposure time of 240 sec per filter. A 
four point square dither pattern was used to assist with the identification and 
rejection of bad pixels and cosmic rays, as well as with the construction of 
on-sky calibration frames. The final images have FWHM = 0.7 arcsec.

	The data reduction sequence for each exposure consisted of (1) the 
subtraction of a dark frame, (2) the division by a dome flat, which 
was obtained by differencing images of a dome spot recorded with the 
lights on and off, (3) the subtraction of the 
DC sky level, and (4) the subtraction of 
interference fringes and the thermal signatures of objects along the 
optical path, using a calibration frame that was constructed by 
median-combining flat-fielded and sky-subtracted images of various fields.
The processed images were aligned to correct for the dither
offsets, median-combined, and then trimmed to the region having a full 240 sec 
exposure time. The final $K'$ image of NGC 205 is shown in Figure 1.

	Stellar brightnesses were measured with the point-spread function 
(PSF) fitting program ALLSTAR (Stetson \& Harris 1988), using target lists, 
preliminary photometric measurements, and PSFs obtained from tasks in 
DAOPHOT (Stetson 1987). The photometric calibration was based on 
observations of UKIRT faint standard stars (Hawarden et al. 2001). 
Completeness and the photometric uncertainties due to 
crowding and sky noise were estimated from artificial star experiments, 
and the results are summarized in Figure 2 for the two radial intervals 
used in the photometric analysis (\S 3). Incompleteness 
becomes significant when $J$ and $K'$ are fainter than 18th magnitude.

\section{COLOR-MAGNITUDE DIAGRAMS AND COLOR DISTRIBUTIONS}

	The CFHTIR field was divided into two equal area regions 
centered on the galaxy nucleus to investigate radial 
trends in stellar content. The `inner' region has 
a radius of 80 arcsec, while the `outer' region covers the remainder of the 
field. Based on the $r-$band surface photometry measurements made by Kent 
(1987), the inner region has an integrated brightness M$_r = -15$, 
while for the outer region M$_r = -14$.

\subsection{Color-Magnitude Diagrams and the Incidence of Blending}

	The $(K, H-K)$ and $(K, J-K)$ color-magnitude diagrams (CMDs) of the 
inner and outer regions are shown in Figure 3. Ferraro et al. (2000) calibrated 
the $K-$band RGB-tip brightness in globular clusters as a function of 
metallicity, and their relation predicts that M$_{K}^{RGBT} = -6.5$ 
for an old population with [Fe/H] $= -0.85$, which corresponds roughly to 
$K = 18$ at the distance of NGC 205. The artificial star 
experiments indicate that incompleteness and errors in the photometry become 
significant when $K \geq 18$, while the incidence of blending between stars 
will also increase markedly when $K \geq 18$ due to the onset of the RGB. 
Therefore, the present study focuses on stars with 
K $\leq 18$, which are evolving on the upper AGB. 

	The AGB in the CMDs has two components: (1) a vertical 
sequence, which consists of oxygen-rich K and M 
giants, and (2) a red plume with $K < 17.2$ and $H-K > 0.4$ 
and $J-K > 1.5$, which contains C stars (e.g. Hughes \& Wood 1990; Wood et al. 
1985). The upper envelopes of sources in the inner and outer regions differ by 
0.4 magnitudes, in the sense that the brightest sources occur in the inner 
region. The stellar density in the inner region is higher than in the outer 
region, and this raises the concern that the brightest stars in the inner 
region may be blends. The spatial distribution of the brightest stars 
in the inner region suggests that this is not the case. If these stars 
were blends, then they would be concentrated near the nucleus. However, the 
stars with $K < 16$ are scattered throughout the inner region, including along 
the minor axis near the edge of the inner region. A more quantitative 
assessment of the incidence of blending than can be infered from the spatial 
distribution of the brightest stars is highly desireable, and this was done in 
the present study using two different approaches, both of which assume that 
there is not a significant population gradient. While there almost 
certainly {\it is} a population gradient in NGC 205 (see below), accounting 
for this gradient would not alter the results of this analysis by 
a significant amount. 

	The probability of blending can be estimated from star counts. However, 
the star counts upon which the estimates are based may in turn be affected 
by blending. One way to lower the chances of this being an issue is to measure 
number counts in a field that has a lower surface brightness 
than that being studied, and then scaling 
the results to match those expected in the higher surface brightness field. 
For the following calculations, the number counts are based on measurements in 
the outer field, and these are scaled to match those expected in the inner 
field to estimate the number of blends.

	If two stars of equal magnitude fall in the same resolution element on 
the sky they will appear as a single source that is 0.75 mag brighter than the 
unblended stars \footnote[2]{The discussion is restricted to blends involving 
two stars, as the incidence of blends between three or more stars 
of comparable brightness is considerably lower than for two stars.}.
The inner region contains stars with $K$ between 15.8 and 16.2 that 
are not seen in the outer field, and if these are the result of blends 
then the unblended stars will have $K$ between 16.6 
and 17.0. There are 130 stars with $K$ between 16.6 and 17.0 
in the outer region, where the mean surface brightness is $\mu_r = 21.3$. 
If the radius of each resolution element is 
one half the FWHM of the PSF (i.e. 0.35 arcsec), then the density of stars with 
$K$ between 16.6 and 17.0 is $2.50 \times 10^{-3}$ per resolution element, 
and the expected number of two-star blends in the outer region is then 0.3.
The mean surface brightness in the inner region is $\mu_r = 20.7$, 
and if the stellar content is like that in the outer region then 
there will be $4.34 \times 10^{-3}$ stars per resolution element 
with $K$ between 16.6 and 17.0. The number of sources with $K$ between 
15.8 and 16.2 in the inner regions that are due to two-star blends is then 0.9. 

	Another way to assess the effects of blending is to combine low surface 
brightness fields to simulate regions of high surface brightness. The 
brightnesses of stars in the original and simulated fields can then be compared 
to study directly the effects of crowding. In the current study, seven $100 
\times 100$ pixel sub-fields, located along the minor axis of NGC 205 at the 
edge of the CFHTIR field and with surface brightnesses $\mu_r = 21.6$ mag 
arcsec$^{-2}$ based on the measurements in Table III of Kent (1987), were 
combined to create a field with a surface brightness $\mu_r = 19.5$ mag 
arcsec$^{-2}$, to simulate the $100 \times 100$ 
pixel region centered on the nucleus of NGC 205.

	The brightest source in the simulated high 
density field is 0.1 mag brighter in $K$ than the 
brightest source in the original fields, indicating that the peak 
stellar brightness is not sensitive to stellar density, even near the 
nucleus of NGC 205. While there are 34 sources in the top 2 magnitude 
interval in the original fields, 44 sources were found in this same interval 
in the simulated field, and this difference is due to blends of fainter stars 
in the simulated field that appear as brighter composite objects. These 
simulations indicate that blending is a concern only well below the 
AGB-tip; even in the densest regions of the inner field the majority 
of objects within two magnitudes of the AGB-tip are unblended stars.
It should be emphasized that the mean surface brightness in the inner field 
is 1.2 magnitudes arcsec$^{-2}$ lower than in the $100 \times 100$ pixel 
field simulated here, so the effects of blending throughout most of the 
inner field are less than in this worse-case simulation.

\subsection{Color Distributions}

	The histogram distributions of the $H-K$ and $J-K$ colors of stars with 
$K$ between 16.2 and 17.2 are shown in Figures 4 and 5. The color distributions 
have a main peak, containing K and M giants, and a red tail, containing C 
stars. Also shown are Gaussians with $\sigma$'s equal to the random photometric 
errors predicted from the artificial star experiments, and normalized to match 
the number of stars in the central bins of the M giant peaks. C stars produce 
an excess population of objects with respect to K and M giants when $H-K > 0.4$ 
and $J-K > 1.5$ in both regions. The M giant sequence in the inner region also 
has a blue tail when $J-K < 1.2$, which is due to younger AGB stars 
than those along the M giant peak. A corresponding 
blue tail is not seen in the outer region, indicating that the youngest 
AGB stars are restricted to the inner region. This is qualitatively consistent 
with the difference in peak brightness between the inner and outer regions.

	The AGB becomes bluer towards fainter 
magnitudes, and if fainter stars occur in sufficient numbers to blend 
together, they will appear as brighter objects 
with bluer colors than the main AGB locus. Could the blue population 
of stars in the inner region $J-K$ color distribution be the result 
of blends? This question is answered using the procedure described in \S 3.1. 
The blue population in Figure 5 has a mean color $J - K = 
1.1$ and, if due to blends, would come from stars that are 0.8 mag 
fainter, and having similar colors. The color distributions in Figure 5
are based on stars with $K$ between 16.2 and 17.2, and 
there are 123 stars in this magnitude range with $J-K$ between 1.0 and 
1.2. For comparison, there are 280 stars with $K$ between 17.2 
and 18.2 and $J-K$ between 1.0 and 1.2 in the outer region. 
After scaling upwards to match the surface brightness in the inner field, 
the density of stars with $K$ between 17.2 and 18.2 expected in the inner 
region is then 0.0093 stars per resolution element. These will produce $4 - 5$ 
blends with $K$ between 16.2 and 17.2 in the inner region. 
The number of suspected blends is thus much smaller than the number of detected 
stars in the blue envelope, indicating that this feature 
is not an artifact of blending.

\subsection{Comparison with Models}

	Girardi et al. (2002) compiled isochrones in the $JHK$ photometric 
system, and these have been compared with the peak 
stellar brightnesses and the $J-K$ color distribution of the brightest 
stars in each region. The transformed models, which include evolution on the 
thermally-pulsing AGB, do not apply to C stars. The comparisons are restricted 
to two metallicities: z = 0.004, which corresponds roughly to 
the RGB metallicity measured by Mould et al. (1984), and
z = 0.019, which was considered since the brightest AGB stars are younger than 
the main body of RGB stars, and so might be more metal-rich. Caution should 
of course be excercised when making comparisons between computed and 
observed quantities, as there are uncertainties in the transformation onto the 
observational plane and in the input physics, due to uncertainties in the 
core-mass versus luminosity relation, and the methods used to model convection, 
mass loss, and envelope burning.

	The predicted relations between AGB-tip brightness and age for z=0.004 
and z=0.019 are shown in Figure 6. The peak $K$ brightness of stars 
in the inner region suggest that the youngest stars have either 
log(t$_{yr}) \sim$ 8.0 or 8.6, with the colors of these objects favouring 
the lower value (see below). As for the outer region, the peak $K$ brightness 
is consistent with the youngest stars having log(t$_{yr}) < 9.0$. 

	The models in Figure 6 predict that the peak AGB brightness in a 
population with log(t$_{yr}$) = 10.0 should occur 
near $K = 17$ at the distance of NGC 205, and so 
the color distributions in Figure 5 are dominated by stars that formed during 
intermediate epochs. The comparisons with the predicted colors in Figure 5 
indicate that the mean $J-K$ color of the blue tail in the inner region 
color distribution is consistent with these stars having  
log(t$_{yr}) \sim 7.8$, which agrees with the lower age estimate predicted from 
the peak AGB brightness in Figure 6. The models also indicate that the 
stars in the blue tail in the top panel of Figure 5 are at least 
a few tenths in log(t$_{yr}$) younger than the stars that contribute to 
the main peak in the color distribution. Finally, the model predictions 
in Figure 5 also demonstrate that the ability to resolve age differences 
diminishes with increasing age, and the M giant peak in 
the color distributions could contain stars spanning a range of ages.

	If there is an age-metallicity relation among stars in NGC 205 
then this will also cause a spread in the colors of upper AGB stars over 
and above that expected only from age effects. However, 
only a modest spread in metallicity is likely present. 
At a given age, the z = 0.004 and z = 0.019 models have $J-K$ colors 
that differ by 0.1 magnitude on the upper AGB, indicating that 
$\frac{\Delta [M/H]}{\Delta J-K} = 7.0$ dex mag$^{-1}$. For comparison, 
photometric errors introduce a $\pm 0.1$ magnitude spread in $J-K$ on the upper 
AGB. Given that the width of the K and M giant peak in the $J-K$ color 
distribution in Figure 5 is comparable to that predicted by photometric 
errors then if a spread in metallicity is present, it can only introduce a 
spread of a few hundredths of a magnitude in $J-K$, which in turn corresponds 
to $\Delta$[M/H] = 0.1 -- 0.2 dex.

\section{THE LUMINOSITY FUNCTIONS OF M GIANTS AND CARBON STARS}

	Bolometric corrections in the $K-$band, BC$_K$, were computed for 
individual stars using the relation between BC$_K$ and $J-K$ for Galactic 
and LMC AGB stars given by Bessell \& Wood (1984). While this relation is 
based on oxygen-rich stars, it is evident from Figures 3 and 4 of 
Bessell \& Wood (1984) that it gives BC$_K$s for C 
stars that are reliable to within a few tenths of a magnitude.

	The composite bolometric LFs of oxygen and 
carbon-rich giants in the inner and outer regions are compared in 
Figure 7. Objects with M$_K < -9.5$ and/or M$_{bol} > -7$ were not considered 
to be AGB stars, as these are likely red supergiants or star clusters. 
After removing these objects, the brightest star in the CFHTIR field has 
M$_{bol} = -6.5$, while the next two brightest stars have M$_{bol} \sim -6.4$. 
These three stars fall along the M giant sequence, and hence are likely not 
C stars. The LFs in Figure 7 were not extended fainter than M$_{bol} = -3.75$ 
to avoid (1) sample incompleteness in excess of 50\%, which occurs when 
M$_{bol} \geq -3.7$, and (2) the RGB-tip, which occurs near M$_{bol} = -3.6$ in 
moderately metal-poor populations (e.g. Ferraro et al. 2000).

	The LFs in the upper panel of Figure 7 are not parallel, indicating 
that the inner and outer regions have different AGB contents. To better 
compare these data, the LF of the outer region was scaled upwards to 
match the number density of stars in the inner region based on 
the $r-$band surface brightness measurements made by 
Kent (1987), and the result is shown in the lower panel of Figure 7. When 
compared in this manner, the LFs of the two regions differ 
significantly between M$_{bol} = -5.75$ and $-4.75$, in the sense that the 
inner region contains a higher fraction of 
luminous AGB stars. While the LFs do not differ significantly 
when M$_{bol} < 5.75$, the LF of the inner region still falls above 
that of the outer region in this interval. These results are consistent with 
the youngest stars being concentrated towards smaller radii.

	The comparisons in the lower panel of Figure 7 provide further 
evidence that blending is not a significant problem among bright AGB stars in 
the CFHTIR field. If a significant amount of blending occurs in 
the inner region, then this will be most evident near the faint end of the AGB 
LF, which terminates just above the RGB-tip, in the sense that the number of 
sources in the inner region will be higher than in the outer region after 
accounting for differences in stellar density using surface brightness 
measurements. The LFs in the lower panel of Figure 7 are in excellent 
agreement at the faint end, as expected if blending is not a significant 
issue in the inner region.

	Davidge (1992) examined the AGB content in a 2.2 square arcmin field 
centered on the nucleus of NGC 205, and in the lower panel of Figure 7 the 
composite LF from that study, scaled to match the number of stars in the inner 
region between M$_{bol} = -5.5$ and --4.5, is shown. The Davidge (1992) LF 
matches that of the inner region at the faint end, but falls above the 
inner region LF when M$_{bol} = -5.5$. The Davidge (1992) observations sample 
a smaller range of galactocentric radii than is covered in the inner region, 
and the comparison in the lower panel of Figure 7 is thus consistent with 
the number density of stars with M$_{bol} = -5.5$ increasing 
towards the nucleus of NGC 205.

	The bolometric LF of only C stars, which are assumed to 
have $J-K > 1.5$ and $K < 17.2$, is shown in Figure 8. The 
$J-K = 1.5$ criterion is based on the $2-\sigma$ width of the M giant AGB 
sequence at $K = 17$ in Figure 5, and is only slightly different from the 
$J-K = 1.6$ criterion adopted by Hughes \& Wood (1990). 
The $K < 17.2$ selection criterion is based on the approximate lower 
envelope of the C star plume in the CMDs, and this brightness cutoff 
largely limits the C star sample to M$_{bol} < -4.1$.

	Richer et al. (1984) used narrow-band images that monitored the 
strength of CN and TiO absorption bands to identify seven C stars near the 
center of NGC 205. While all seven of these stars are in the CFHTIR 
field, only four can be identified with confidence 
in the CFHTIR images\footnote[3]{C stars \# 1, 5, and 6 
are located close to the center of NGC 205, and can not be unambiguously 
identified using the finding chart in Figure 7 of Richer et al. (1984)}, 
and the colors and brightnesses of these are listed in Table 1. 
The near-infrared colors and brightnesses of these stars place them 
on the red plume on the $(K, H-K)$ and $(K, J-K)$ CMDs in Figure 3, and 
these objects are successfully identified as C stars
using the brightness and color criteria described in the previous paragraph.

	The brightest C stars likely 
formed in the most recent star-forming episode, and while the LFs of both 
regions in the lower panel of Figure 8 are similar when M$_{bol} > -5$ 
after adjusting for differences in surface brightness, the 
outer region has a deficiency of stars with M$_{bol} = -5.5$ when compared 
with the inner region. This is consistent with the comparison 
between the inner and outer region LFs in Figure 7. 

	The large number of C stars in the CFHTIR field is worth noting; 
after adjusting for sample incompleteness, 224 C stars with 
M$_{bol} < -4.1$ are found in the inner field, while 96 are found in 
the outer field. This is not a complete census of C stars near the center 
of NGC 205, as `warm' C stars with $J-K < 1.5$, which may account for $\sim 
20\%$ of the sources with $J - K < 1.5$ in the LMC (e.g. Wood et al. 1985), are 
missed. Nevertheless, C stars still account for a significant fraction of the 
total light from AGB stars near the center of NGC 205. The 
integrated bolometric magnitude of C stars with $J - K > 1.5$ is M$_{bol} = 
-10.6$ in the inner region, and --9.6 in the outer region. For comparison, 
the integrated bolometric magnitude of all oxygen and carbon-rich AGB 
stars with M$_{bol} < -3.75$ is --13.0 in the inner region, and --12.2 in the 
outer region. C stars thus contribute at least 11\% of the integrated AGB 
luminosity in the inner region, and 9.1\% in the outer region.
C stars make a similar contribution to the integrated AGB luminosity of LMC 
clusters with ages between 0.4 and 2 Gyr (Maraston 1998).

\section{DISCUSSION AND SUMMARY}

	$J, H,$ and $K'$ images with sub-arcsec angular resolution have been 
used to investigate the near-infrared photometric properties of bright AGB 
stars in the Local Group dE galaxy NGC 205. The 
$(K, H-K)$ and $(K, J-K)$ CMDs of NGC 205 split into two branches near 
the bright end, with a vertical sequence made up of 
oxygen-rich K and M-type giants, and a red plume containing C stars. The 
onset of the C star plume occurs at $H - K = 0.5$ and $J - K = 1.5$, which 
is roughly consistent with the near-infrared properties of cool C stars in the 
LMC (Hughes \& Wood 1990; Wood et al. 1985). 

	The data suggest that the M-giant AGB sequence in NGC 205 terminates 
near M$_{bol} = -6.5$, while the most luminous C stars have M$_{bol} = -5.5$. 
A source of uncertainty in these values is that the 
brightest AGB stars are very rare, and small number statistics 
may cause the brightness of the AGB-tip to be underestimated, as was found to 
be the case in NICMOS observations of the outer regions of NGC 
5128 (Davidge 2002). In addition, the brightest 
AGB stars may be variable, and this will introduce 
uncertainies of a few tenths of a magnitude in the peak AGB brightness. 
In fact, if the measured AGB-tip brightness is based on
LPVs at the peak of their light curves, then this will 
bias upwards the luminosity of the AGB-tip. The effects of variability 
on the luminosity of the AGB-tip can be checked by re-observing this field.

	The uncertainties in the luminosity of the AGB tip notwithstanding, 
C stars contribute at least 10\% of the total luminosity coming from all 
AGB stars with M$_{bol} < -3.75$ near the center of NGC 205. This is a lower 
limit, as warm C stars with $J-K < 1.5$ are not identified in 
the current census. Therefore, based on the fuel consumption theorum, 
it can be concluded that {\it at least} 10\% of the nuclear fuel consumed 
during AGB evolution near the center of NGC 205 is processed by C stars.
This is consistent with what is seen in intermediate age 
clusters in the LMC (Maraston 1998).

	The inner region of NGC 205 contains an excess population of AGB stars 
with respect to the outer region when M$_{bol} < -4.75$, indicating that 
the brightest AGB stars are not uniformly mixed with fainter stars 
near the center of NGC 205; rather, there is an age gradient. 
Based on the peak brightness of AGB stars 
in the inner field, and the $J-K$ colors of the blue AGB sequence in Figure 5, 
the Girardi et al. (2002) isochrones indicate that the youngest 
evolved stars near the center of NGC 205 have ages log(t$_{yr}) < 8.0$. 
This is consistent with the ages of star clusters near 
the center of NGC 205 measured by Cappellari et al. (1999). 

	Cepa \& Beckman (1988) investigated the orbit of NGC 205 about M31, and 
concluded that (1) the orbital period is 0.3 Gyr, and (2) 
NGC 205 last crossed the disk of M31 0.1 Gyr in the past. 
The colors and brightnesses of the youngest AGB 
stars in the inner region, which include the brightest 
AGB stars and the blue tail in the $J-K$ distribution, are consistent with 
these objects having formed during the most recent crossing of the M31 disk, 
suggesting that this interaction likely spurred star formation in NGC 205. 
The relative amplitudes of the blue and red 
peaks in the inner region $J-K$ color function indicates that the older M 
giant sequence contains 3 times more stars than formed during the most recent 
interaction. The color offsets between the blue AGB stars in the inner region 
and the older M giant peak in Figure 5 is consistent with a hiatus of at least 
a few tenths of a Gyr between the most recent and any previous star-forming 
episode, once again in agreement with the Cepa \& Beckman (1988) orbital 
parameters. Interactions with NGC 205 have evidently not had a major impact on 
the star formation rate in the disk of M31, which has been low outside of 
spiral arms for the past 1 Gyr (Williams 2002).

	We close by noting that the observations used in this paper amount to 
only a few minutes total integration time per filter on a 3.6 metre telescope, 
and the resulting data are able to clearly separate 
the C star and M giant sequences on near-infrared 
CMDs. $J, H,$ and $K'$ images thus provide an efficient means of identifying 
C stars in nearby galaxies. With adaptive optics (AO) systems on large 
telescopes it should be possible to probe the AGB content of galaxies 
outside of the Local Group. That a C star plume is clearly seen in the $(K, 
H-K)$ CMD of NGC 205, indicates that observations in $H$ and $K$, where 
AO systems not intended for use in the thermal infrared regime 
will deliver the highest Strehl ratios, should be 
sufficient to detect C stars. This being said, it is worth noting 
that C star sequences are not seen in the $(K, H-K)$ CMDs 
of the central regions of M32 (Davidge et al. 2000) and M31 (Davidge 2001). If 
there is an inverse correlation between C star content and metallicity, as 
suggested by Brewer, Richer, \& Crabtree (1996), then the absence of C 
stars in M32 and the bulge of M31 is likely due to the relatively high 
metallicities of these systems, although it is still not clear if the 
central regions of these galaxies contain stars as young as those in NGC 205.

\acknowledgements{It is a pleasure to thank an anonymous referee for comments 
that greatly improved the paper.}

\clearpage
\begin{center}
FIGURE CAPTIONS
\end{center}

\figcaption[f1.eps]
{The final $K'$ image of NGC 205, which covers $3.5 \times 3.5$ 
arcmin. Stars in this image have FWHM = 0.7 arcsec. North is at the top, 
and east is to the left. The circle marks the boundary between the 
inner and outer regions used in the photometric analysis.}

\figcaption[f2.eps]
{The results from the artificial star experiments for the two radial intervals 
used in the photometric analysis. The results for the $K-$band are shown 
as solid lines, while the results for the $J-$band are shown as dashed curves. 
The completeness is the number of recovered artificial stars divided by the 
total number that were added, while $\Delta$M is the mean difference in 
magnitudes between the input and measured brightnesses, and $\sigma$ 
is the standard deviation about $\Delta$M. $\Delta$M increases towards 
fainter magnitudes because fainter stars are more easily detected 
when they fall on positive noise spikes, which makes them appear brighter.}

\figcaption[f3.eps]
{The $(K, H-K)$ and $(K, J-K)$ CMDs of the inner (top row) and outer 
(bottom row) regions.} 

\figcaption[f4.eps]
{The histogram distribution of the $H-K$ colors of stars with $K$ 
between 16.2 and 17.2 in the inner and outer regions. The solid line shows 
a gaussian with a standard deviation 
equal to the random photometric errors predicted from the 
artificial star experiments, and scaled to match the number of points in the 
central bins of each distribution. A red tail, due to 
C stars, is seen in both regions when $H-K > 0.4$.}

\figcaption[f5.eps]
{The histogram distribution of the $J-K$ colors of stars with $K$ 
between 16.2 and 17.2 in the inner and outer regions. The solid line shows 
a gaussian with a standard deviation equal to the random 
photometric errors predicted from the artificial star experiments, 
and scaled to match the number of points in the central bins of each 
distribution. A red tail, due to C stars, is seen in both regions when 
$J-K > 1.5$. An excess population of stars when $J-K < 1.2$ is seen in the 
inner region, but not in the outer region. The 
predicted colors for three z=0.019 isochrones from 
Girardi et al. (2002), assuming $\mu_0 = 24.6$ and E(B--V) = 0.035 are 
also shown. The corresponding sequence for z = 0.004, which is not shown here 
to prevent cluttering the figure, is offset by roughly 0.1 mag to smaller $J-K$ 
colors. These models indicate that the stars forming the blue tail in the inner 
region have log(t$_{yr}$) that is at least a few tenths of a dex smaller than 
the stars in the M giant peak. It is also evident that the M giant peak 
may contain stars spanning a wide range of ages.}

\figcaption[f6.eps]
{The relation between the $K$ brightness of the AGB-tip and age, as predicted 
by the Girardi et al. (2002) isochrones for z = 0.019 (solid line) and z = 
0.004 (dashed line). A distance modulus of $\mu_0 = 24.6$ and E(B--V) = 0.11 
have been assumed for NGC 205. The measured peak $K$ brightnesses in the inner 
and outer regions are indicated.}

\figcaption[f7.eps]
{The completeness-corrected composite bolometric LFs of 
oxygen and carbon-rich giants in the inner (solid line) and outer 
(dashed line) regions. A distance modulus of $\mu_0 = 24.6$ and E(B--V) = 0.035 
(Burstein \& Heiles 1982) have been assumed. N$_{0.5}$ is the number of stars 
per 0.5 magnitude interval in M$_{bol}$, and the error bars show the 
uncertainties due to counting statistics and the completeness corrections. 
The observed LFs are compared in the top panel, while in the lower panel 
the LF of the outer region has been scaled upwards to match the mean surface 
brightness of the inner field based on the $r-$band measurements made by Kent 
(1987). The LFs of the inner and outer regions differ between M$_{bol} = 
-4.75$ and --5.75, in the sense that the inner region contains a larger 
fraction of AGB stars per unit $r-$band surface brightness when M$_{bol} < 
-4.75$. The AGB LF from Davidge (1992), which was computed from $V$ and 
$I$ CCD photometry, and has been scaled to match the number of sources in the 
outer field between M$_{bol} = -5.75$ and --4.25, is shown as a dotted line in 
the lower panel. Note the generally good agreement with the 
LFs constructed from the CFHTIR data.}

\figcaption[f8.eps]
{The bolometric LFs of C stars in the 
inner (solid line) and outer (dashed line) regions, corrected for 
incompleteness. A distance modulus of $\mu_0 = 24.6$ 
and E(B--V) = 0.035 (Burstein \& Heiles 1982) have been assumed. 
N$_{0.5}$ is the number of stars per 0.5 magnitude interval in 
M$_{bol}$, and the error bars show the uncertainties due to counting 
statistics and the completeness corrections. The observed LFs 
are compared in the top panel, while in the lower panel the LF of the 
outer region has been scaled to match the mean surface brightness 
of the inner field based on the $r-$band measurements made by Kent (1987). 
The LFs of the two regions differ when M$_{bol} = -5.5$, in the 
sense that the inner region contains a larger fraction of the brightest stars 
per unit $r-$band surface brightness.}

\clearpage

\begin{table*}
\begin{center}
\begin{tabular}{cccc}
\tableline\tableline
\# & $K$ & $J-K$ & $H-K$ \\
\tableline
2 & 16.491 & 1.524 & 0.528 \\
3 & 16.869 & 1.826 & 0.470 \\
4 & 15.795 & 1.652 & 0.605 \\
7 & 16.362 & 1.676 & 0.606 \\
\tableline
\end{tabular}
\end{center}
\caption{Near-Infrared Photometry of C Stars Found by Richer et al. (1984)}
\end{table*}

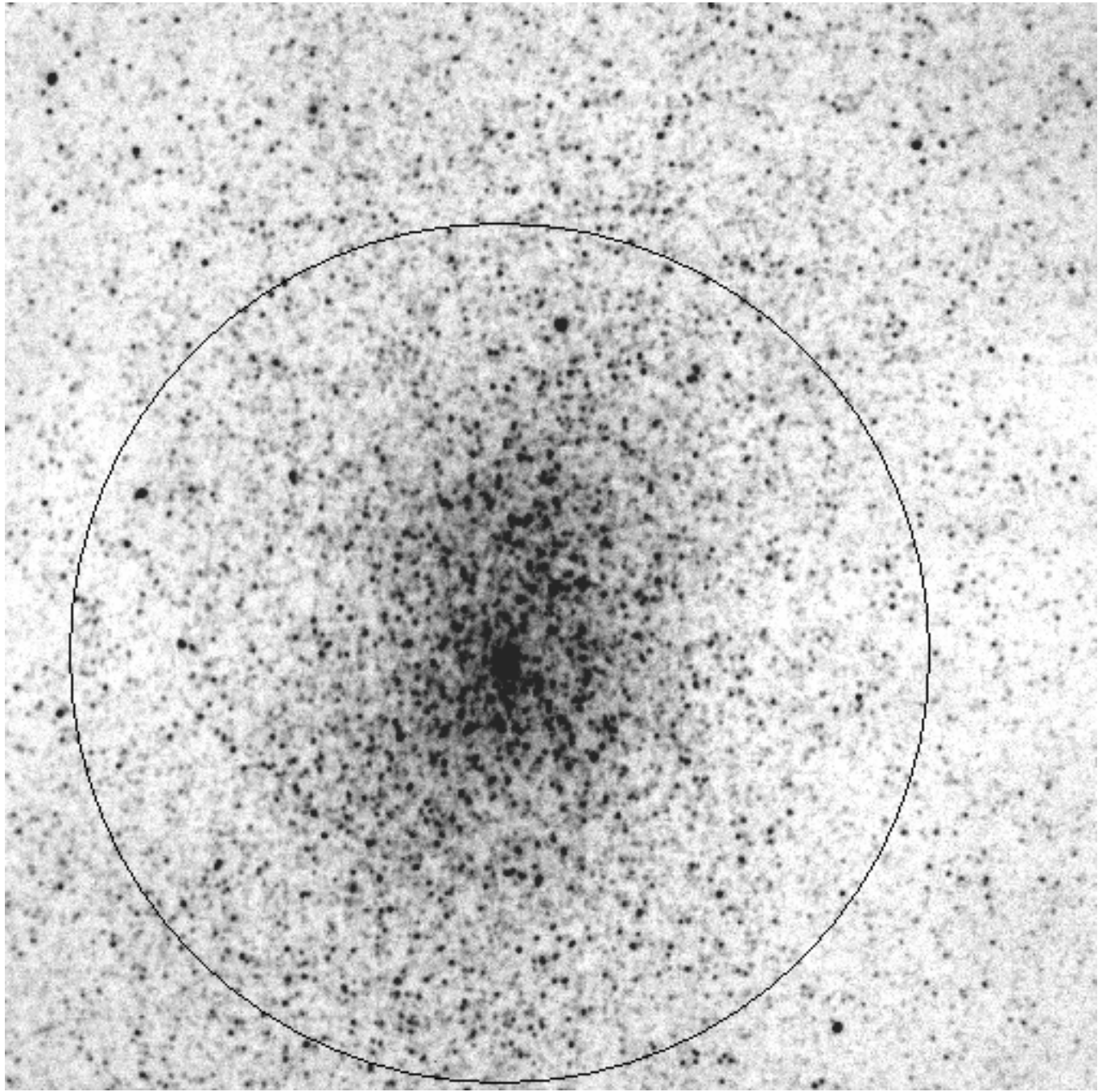
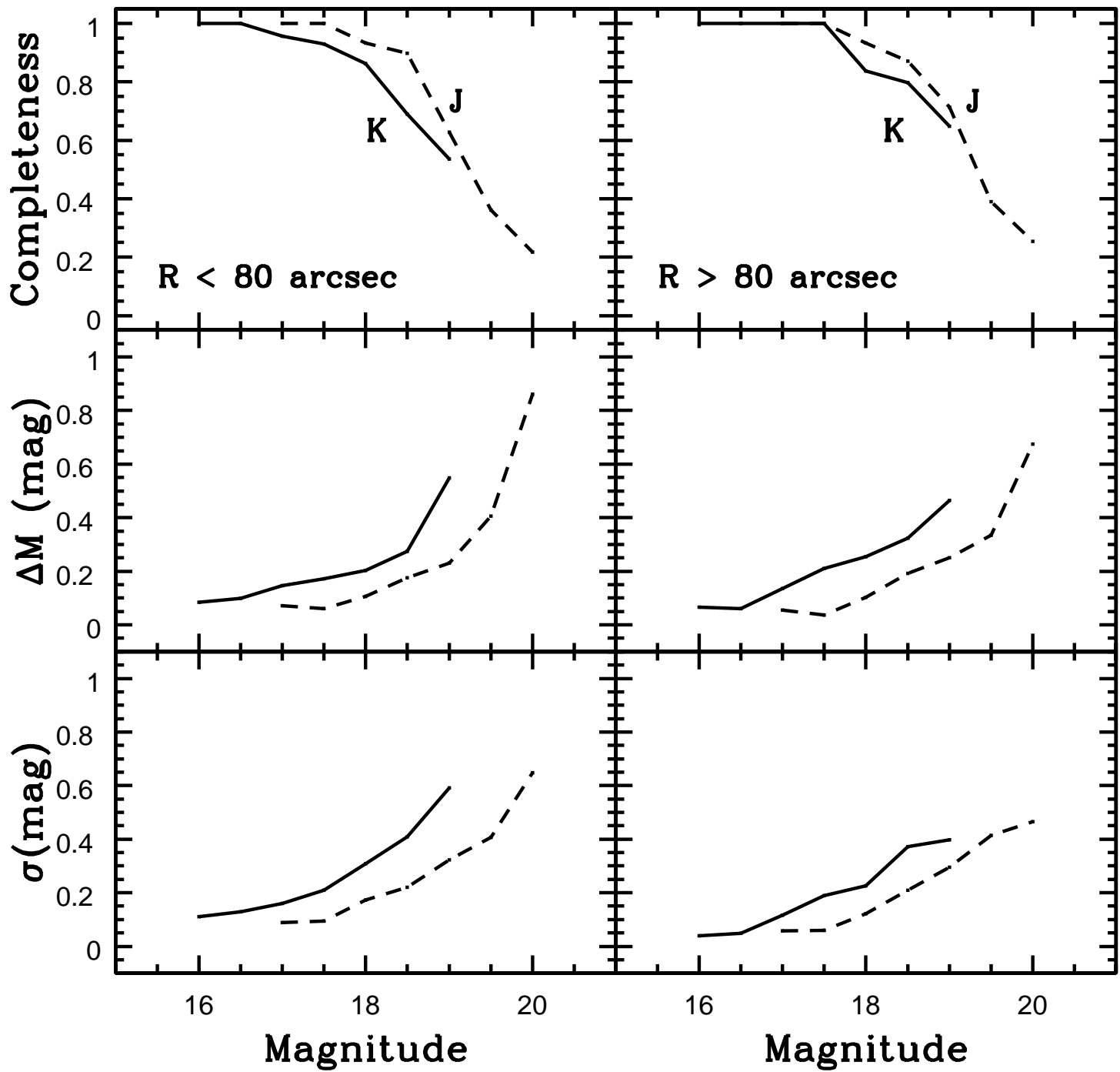
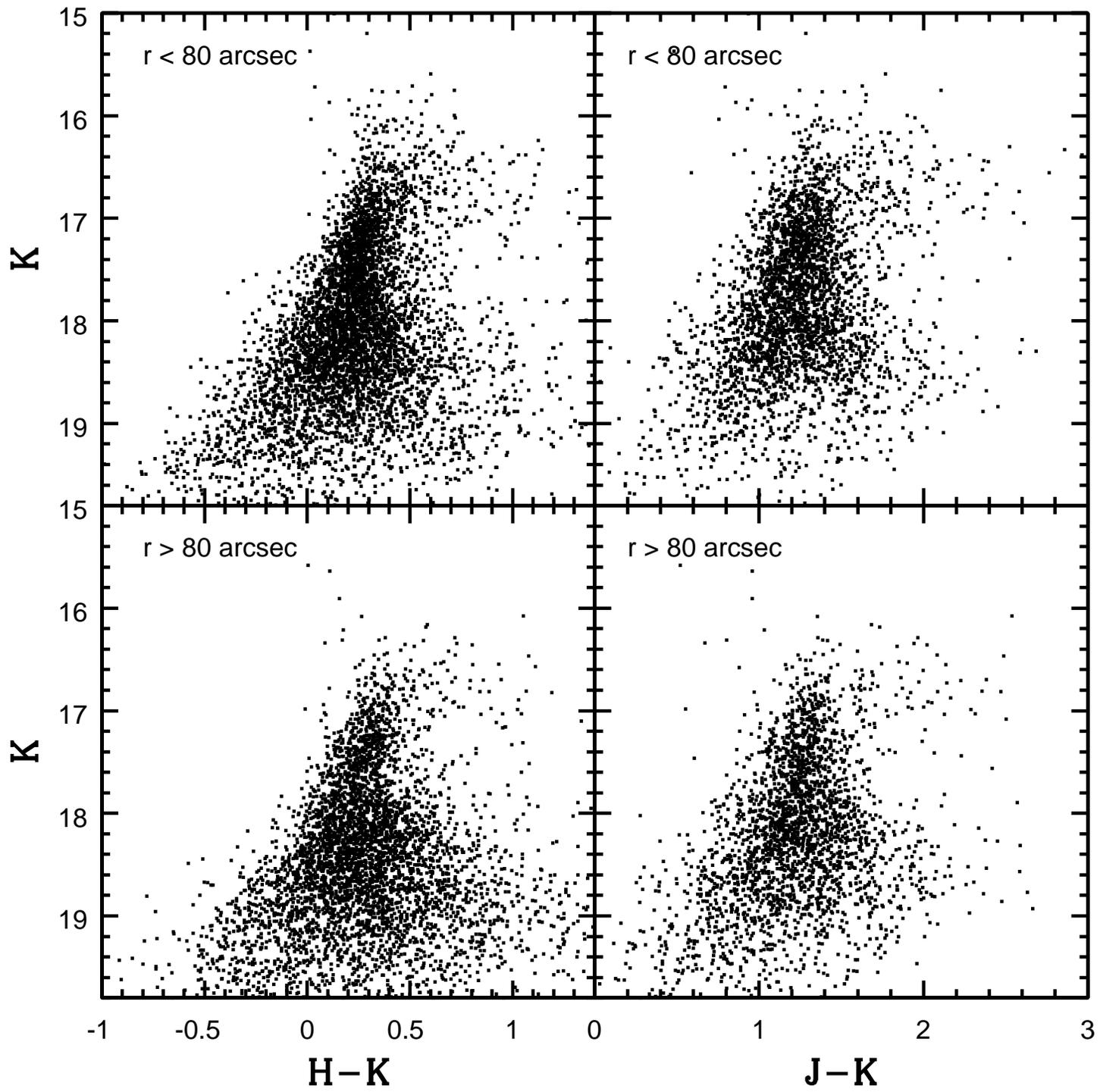
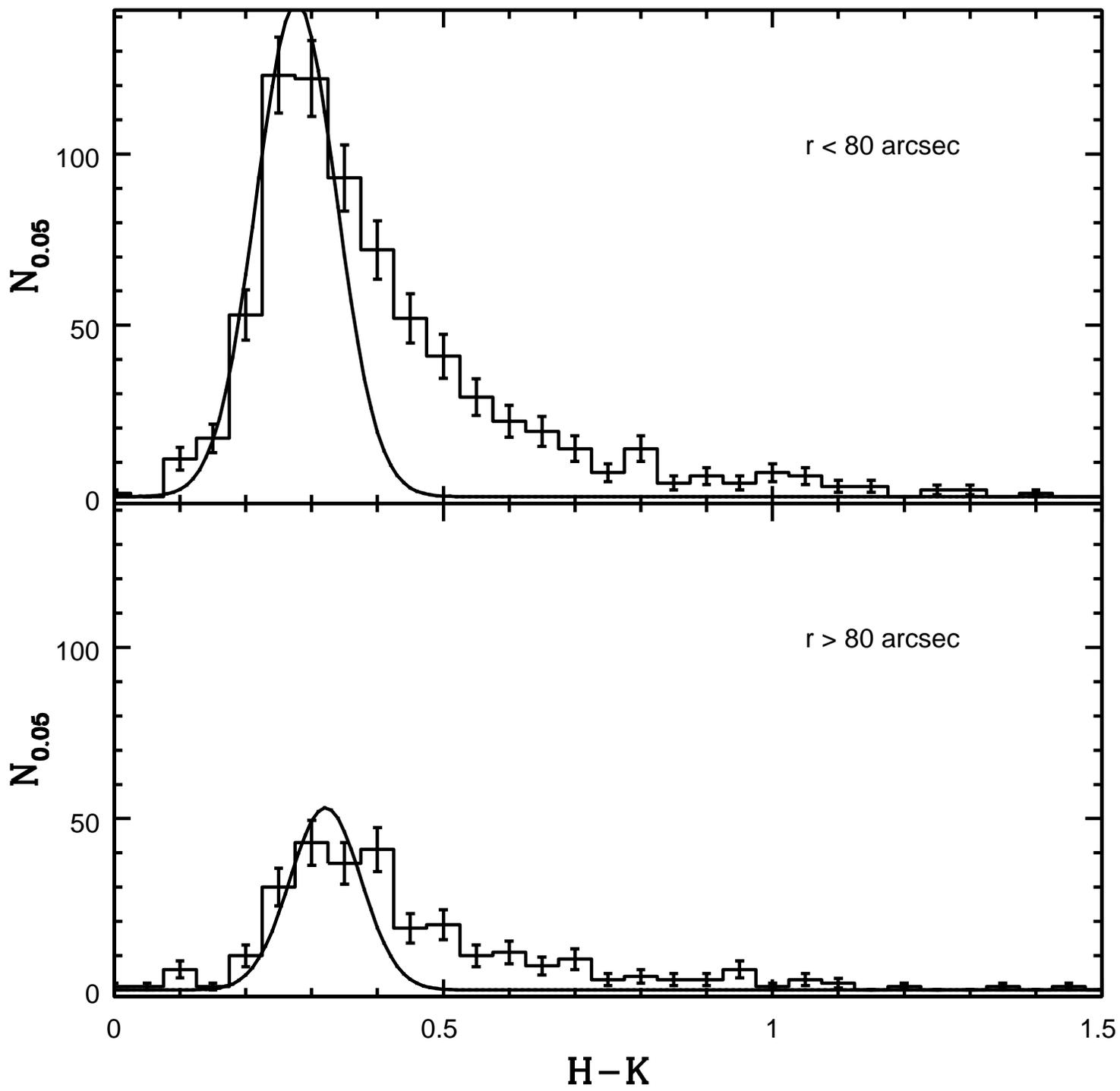
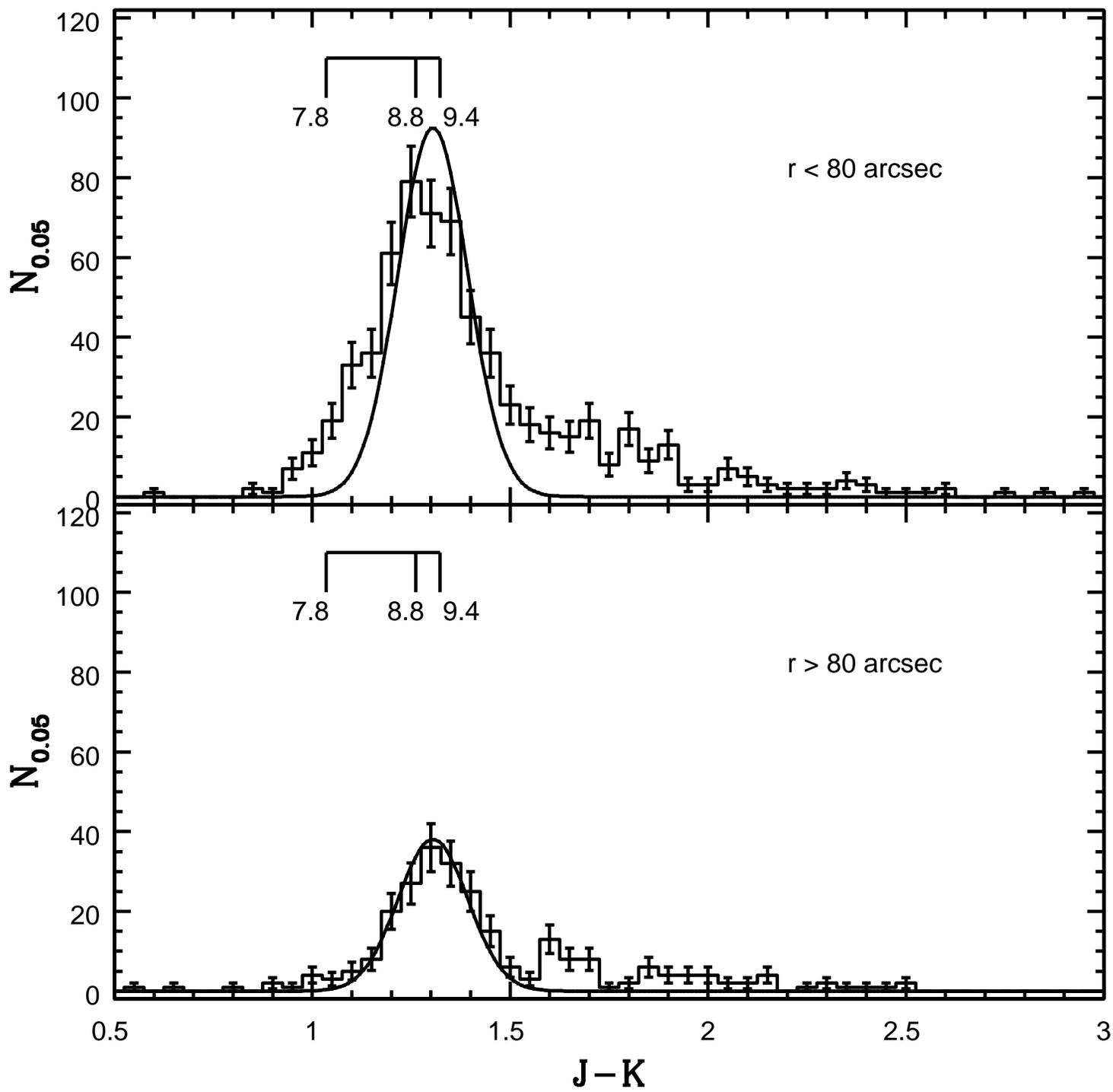
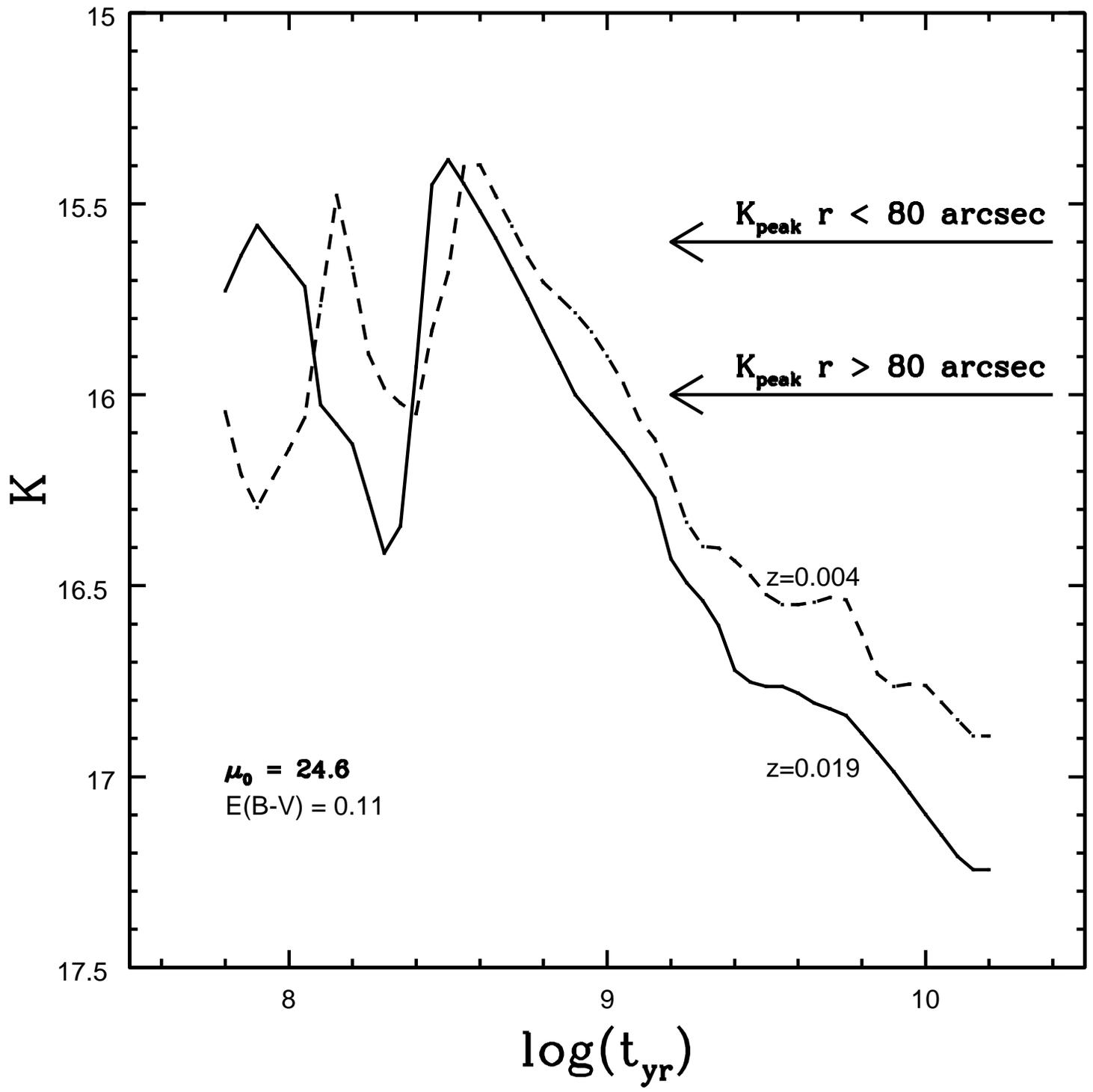
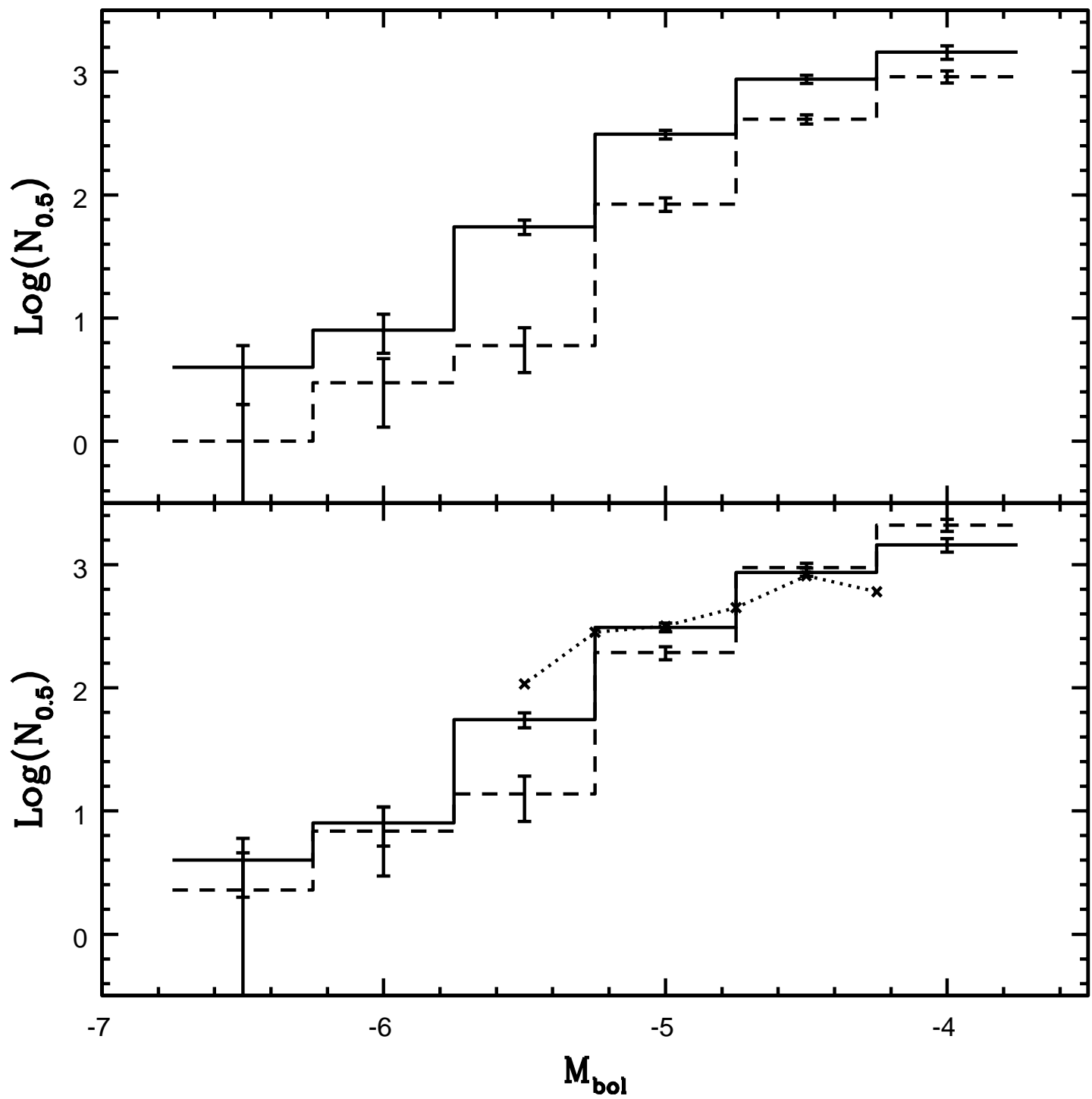
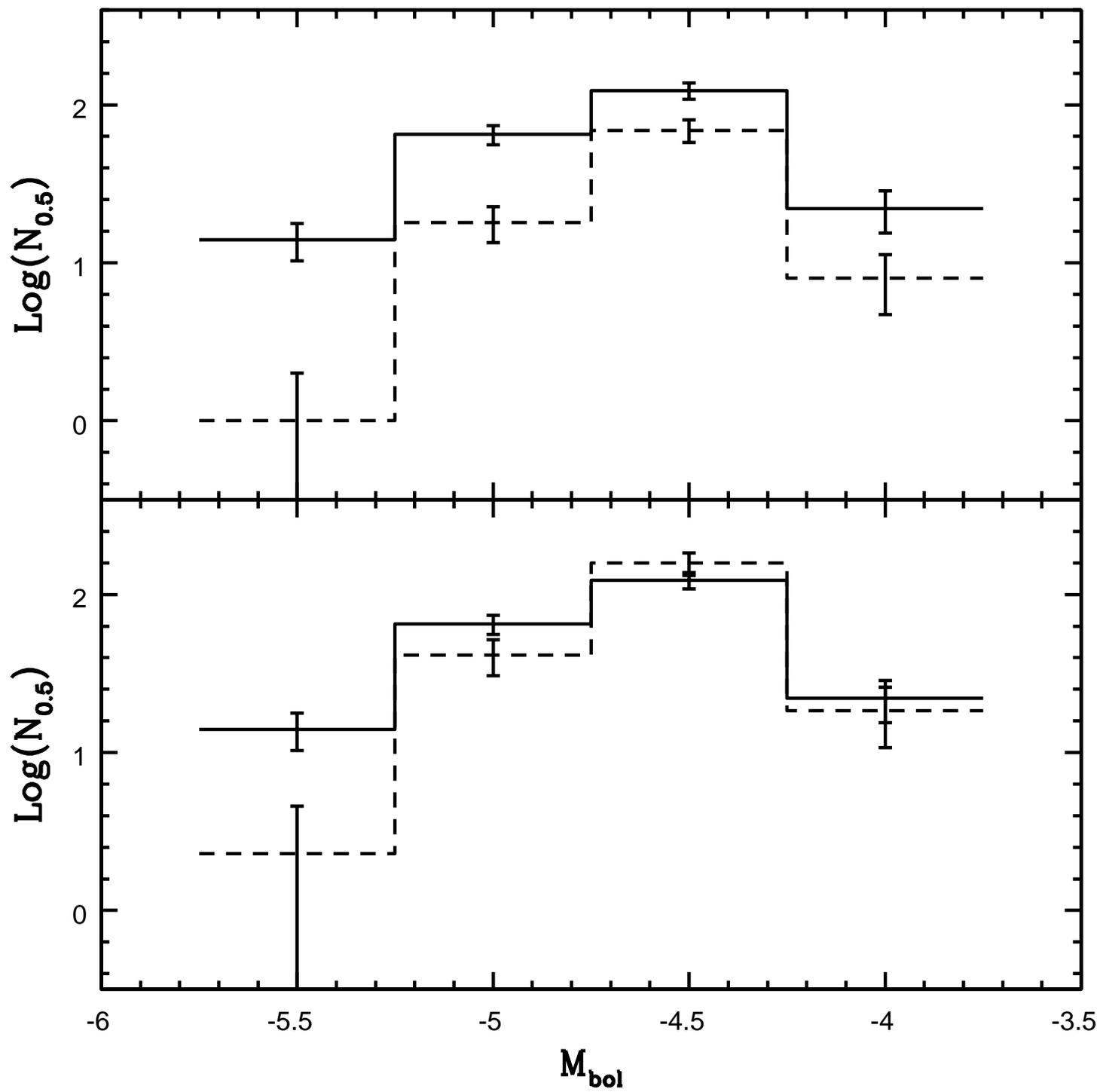
\end{document}